# Quantifying the bulk density of southern delta aquariid meteoroids: insights from the Canadian automated meteor observatory

Arazi Pinhas,[1]★ Zbyszek Krzeminski,[2,3] Denis Vida [2,3] and Peter Brown[2,3]

[1]*Columbus Technologies and Services Inc., Jacobs Space Exploration Group, NASA Marshall Space Flight Center, Huntsville, Alabama 35812, USA*
[2]*Department of Physics and Astronomy, University of Western Ontario, London, Ontario N6A 3K7, Canada*
[3]*Western Institute for Earth and Space Exploration, University of Western Ontario, London, Ontario N6A 5B7, Canada*

## ABSTRACT
Physical properties of ten millimetre-sized meteoroids from the Southern Delta Aquariids (SDA) shower are derived using optical observations from the Canadian Automated Meteor Observatory between 2020 and 2023. The meteors are found to ablate in two distinct erosion stages, the second stage showing a single, bright leading fragment. Our modelling interprets these observations as evidence for equal masses of compact grains embedded in a porous, low density matrix. The average bulk density of SDA meteors is found to be $1420 \pm 100$ kg m$^{-3}$, with the compact component having a density of $2310 \pm 160$ kg m$^{-3}$ and the porous component a density of $700 \pm 110$ kg m$^{-3}$. The high bulk density of SDA meteors is comparable to densities found for the Quadrantid and Geminid showers, both of which also have low perihelion distances. This suggests that thermal desorption may play a significant role in the processing of meteoroids.

**Key words:** comets: general – meteorites, meteors, meteoroids – minor planets, asteroids: general.

## 1 INTRODUCTION

The Southern Delta Aquariids (SDA) meteor shower is a notable celestial event among the primary meteor showers with heightened activity, featuring Zenithal Hourly Rates (ZHRs) of ten or more (American Meteor Society, Ltd 2023). The meteoroid stream is encountered at the ascending node and the shower reaches its annual peak activity on July 30 (Miskotte 2018). The shower exhibits a ZHR of around 20 meteors per hour and a geocentric velocity of ∼41 km s$^{-1}$ (Jenniskens 2006).

The SDA complex has several unusual characteristics. Among these are its very small perihelion distance (∼0.1 au) and long activity period of more than a month (Abedin et al. 2018). The shower is also unusually rich in small particles, being the strongest shower detected by the Advanced Meteor Orbit Radar (AMOR) to a radar limiting magnitude of +14 (Galligan 2000) and the second strongest annual shower detected by the Canadian Meteor Orbit Radar (CMOR) to a radar limiting magnitude of +8 (Brown et al. 2008).

SDA meteoroids show evidence for some of the strongest fragmentation among all shower meteors (Jacchia, Verniani & RE 1967), yet display begin heights that suggest they are materially among the most refractory of all showers. This is demonstrated by their very high values of the $K_B$ proxy strength parameter (Borovička et al. 2019) which studies have found to be either the largest (Joiret & Koschny 2023) or second largest (next to the Geminids) among all major showers (Matlovič et al. 2019). The shower also has a possible linkage with the fragmentation cascade which produced the Kracht sun-skirting comet family (Sekanina & Chodas 2005).

Intriguingly, the shower is most directly associated with comet 96P/Machholz (McIntosh 1990; Abedin et al. 2018), which shows unusual chemistry interpreted to be potentially indicative of an extrasolar origin (Schleicher 2008).

While studies have furnished bulk density estimates for most active meteor showers, the SDA shower has only a couple such measurements. The earliest estimates of SDA bulk densities were based on Super-Schmidt records and reported by Verniani (1967) who found a value for the shower of 300 kg m$^{-3}$ from five reduced meteors in the sub-gram mass range. However, this estimate was based on masses derived assuming single body ablation and using luminous efficiency values now thought to be too low (Popova, Borovička & Campbell-Brown 2019), suggesting the presented bulk density value is an underestimate. Babadzhanov & Kokhirova (2009) reported an average bulk density of $2400 \pm 600$ kg m$^{-3}$ for a set of 8 $\delta$-Aquariid fireballs based on photographic measurements. They applied the quasi-fragmentation model of Babadzhanov (2002) to infer bulk densities based primarily on fits to observed light curves. These fireballs represented meteoroids of gram-sized and larger, but the paper did not specify the distribution between the southern and northern groups within the sample. A few other studies have focused on proxy metrics for material strength or composition of SDA meteors (e.g. Matlovič et al. 2019; Joiret & Koschny 2023). However, such estimates are relative values to other showers using limited information about the atmospheric behaviour of the meteors, and not direct measurements of physical properties such as the bulk density. Consequently, the average bulk density of SDA meteors remains unclear.

Unlike well-studied showers like the Quadrantids (QUA), Lyrids (LYR), Eta Aquariids (ETA), Perseids (PER), Orionids (ORI), Leonids (LEO), and Geminids (GEM) where many density mea-surements across a range of meteoroid masses are available, the SDA shower is comparatively poor in data. Prior research efforts have generated bulk density assessments for QUA (Babadzhanov & Kokhirova 2009; Borovička 2010), LYR (Verniani 1967; Vojáček et al. 2019; Buccongello et al. 2024), ETA (Verniani 1967; Buccongello et al. 2024), PER (Verniani 1967; Bellot Rubio et al. 2002; Babadzhanov 2002; Babadzhanov & Kokhirova 2009; Vojáček et al. 2019; Buccongello et al. 2024), ORI (Verniani 1967; Babadzhanov & Kokhirova 2009; Buccongello et al. 2024; Vida et al. 2024), LEO (Verniani 1967; Babadzhanov 2002; Babadzhanov & Kokhirova 2009; Buccongello et al. 2024), and GEM (Verniani 1967; Bellot Rubio et al. 2002; Babadzhanov 2002; Babadzhanov & Kokhirova 2009; Vojáček et al. 2019; Buccongello et al. 2024). The knowledge gap in the properties of SDA meteors is of practical importance for the scientific community and the National Aeronautics and Space Administration (NASA). Recognized on NASA's monitoring agenda (Moorhead 2023), the SDA shower offers an opportunity to enhance our understanding of meteoroid compositions. The absence of modern bulk density analyses hampers a comprehension of this specific meteor shower, and closing this informational void is important for current and future comparative characterization studies of meteor showers.

★ E-mail: arazi.pinhas@nasa.gov

In this study, we examine optical observations of SDAs using the mirror tracking system of the Canadian Automated Meteor Radar (CAMO). This system provides a temporal cadence of 100 frames per second and a spatial accuracy of the order of 1 m (Weryk et al. 2013; Vida et al. 2021), allowing direct observations of meteor fragmentation and morphology. Employing our implementation of the meteoroid erosion model proposed by Borovička, Spurný & Koten (2007), we infer physical attributes of SDA meteors using a holistic numerical model which is based on forward modelling to reconstruct the observed light curve, deceleration, and wake (Vida et al. 2024). Section 2 outlines the observational equipment and selected meteor events. The ablation model and associated software is outlined in Section 3. We then present results from our forward modelling in Section 4, with a focus on bulk density. We conclude with highlights in Section 5.

## 2 OBSERVATIONS

In this study, we analyse ten SDA meteors observed optically by the CAMO mirror tracking system. Two identical CAMO systems are located about 45 km apart in Southwestern Ontario, Canada. The first system is located near the town of Tavistock (43.26420° N, 80.77209° W) and the second station is near Elginfield (43.19279° N, 81.31565° W). Both systems are oriented northward at a 45° elevation to mitigate interference from moonlight, bright stars, planets, and the Galactic plane. To optimize common meteor observations, the stations' pointings overlap at altitudes between 70 and 120 km. For four events in our study, CAMO data were supplemented with four low-light imaging systems, specifically Electron-Multiplying Charged Couple Device (EMCCD) cameras. The integration of EMCCD cameras improved the model analysis by extending the upper and lower bounds of the light profiles, thereby enhancing confidence in the shape of the light curves. For a comprehensive exploration of EMCCD camera characteristics, readers are directed to the detailed discussion in Gural et al. (2022).

Each CAMO mirror tracking system consists of two cameras. The first is an image intensified widefield camera with a 34° × 34° field of view, which uses the All Sky and Guided Automatic Real-time Detection (ASGARD) real-time meteor detection algorithm (Brown et al. 2010) to identify events with magnitudes brighter than +5.

Within 0.1 s of ASGARD detecting the appearance of a meteor, a mirror pair is cued, redirecting the light of the meteor through an 80 mm telescope with a narrow field of view measuring 1.5° × 1.5°. This telescope is coupled to a second, high 100 frames per second (FPS) image intensified CCD camera with a plate scale of 6 arcsec per pixel (Vida et al. 2021) and a limiting magnitude of about +7. The combination of the telescope and CCD camera provides spatial accuracy down to a few meters and a temporal resolution of 10 ms. This level of spatial and temporal accuracy, together with the mirror tracking capability, allows for detailed observations of meteor fragmentation and measurements of wake.

Manual reduction of optical observations was performed via the methodology outlined in Vida et al. (2021). During this procedure, the light curve, deceleration, and wake profile are measured. Astrometric picks were made by centroiding the meteor's leading fragment in the narrow-field video data, and the photometry was measured by manually selecting all pixels belonging to the meteor, including the wake. Photometry was measured separately using both wide-field and narrow-field data, with the narrow-field data being several magnitudes more sensitive than the wide-field, enabling extended tracking of the meteor's ending. The brightness profile of the wake was measured on a per-frame basis by cropping a rectangular strip encompassing the meteor and its wake from the image, with subsequent subtraction of the background. The wake's profile as a function of distance from the meteor head was estimated as the sum of pixel intensities in every image column.

Fig. 1 illustrates the sample of ten SDA meteors as captured by Elginfield's narrow-field camera. Generally, most events exhibit a smooth tadpole-shaped morphology in the early phase, featuring a bright head and a clearly defined wake (tail). As the trajectory progresses, several fragments become visible, with a leading fragment in front of other fragments which rapidly fall behind.

During modelling, it was found that two separate bulk densities are required to explain the observations – one characterizing the bulk meteor observed during the initial trajectory and another for the latter phase where a dominant main fragment governs the light production. We interpret this to mean that most SDA meteoroids are structurally constructed from two components, the more volatile/lower density material ablating first and the denser, refractory portion, remaining at the end in the form of a dominant leading fragment.

## 3 ABLATION AND FRAGMENTATION MODEL

The analysis of observed meteor events uses the METSIM modelling tool (Vida et al. 2024), a graphical user interface (GUI) for a numerical implementation of the erosion model introduced by Borovička et al. (2007). This model is well-established through its use in prior studies to interpret both faint meteors and meteorite-dropping fireballs (e.g. Borovička et al. 2013; Vojáček et al. 2019; Vida, Brown & Campbell-Brown 2020; Buccongello et al. 2024; Vida et al. 2024).

The erosion model assumes that meteoroids fragment in the atmosphere through the loss of µm-sized constituent grains, each ablating separately as single bodies. Another assumption is that there is no progressive fragmentation, such that the grains do not fragment further once released. The ablation of each grain is computed using the classical single-body equations

$$\frac{\mathrm{d}v}{\mathrm{d}t} = -K m^{-1/3} \rho_a v^2,  \quad (1\mathrm{a})$$

$$\frac{\mathrm{d}m}{\mathrm{d}t} = -K \sigma m^{2/3} \rho_a v^3,  \quad (1\mathrm{b})$$

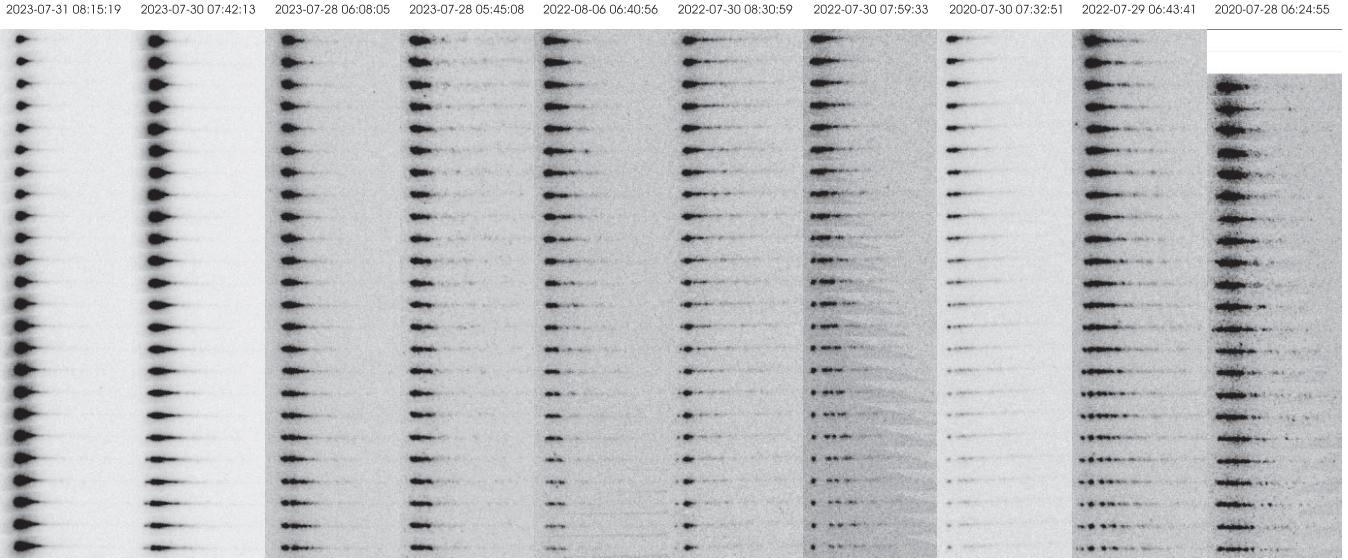

**Figure 1.** Grey-inverted images of ten SDA meteors captured through the CAMO narrow-field camera. The progression of frames from top to bottom follows the passage of time. More than half of the meteors end in a leading meteor that dominates the light output. In all cases the meteor is moving from the right to the left. The dates and UTC time of meteor appearance are given at the top of each column.

$$I = -\tau \frac{v^2}{2} \frac{dm}{dt} + mv \frac{dv}{dt}, \qquad (1c)$$

where $v$ represents the meteor's velocity, $m$ its mass, $\rho_a$ the atmospheric mass density, $\sigma$ the ablation coefficient, $\tau$ the luminous efficiency, and $I$ the luminosity. Following Borovička et al. (2007) all simulated fragments are assumed to be spherical with the shape coefficient of $A = 1.21$ and with a drag factor $\Gamma = 1$. The dynamics are sensitive only to the product of these terms and the bulk density and thus the shape-density coefficient $K = A\Gamma\rho_m^{-2/3}$ is used, where $\rho_m$ is the meteoroid density.

The model involves two types of mass-loss for the main body: ablative mass-loss and erosive mass-loss. The total mass lost into erosion is computed at every time-step using the same single-body equations, but instead of the ablation coefficient a separate erosion coefficient $\eta$ is used. Thus, the effective mass-loss representing the total mass lost by the main body is the sum of the ablation and erosion mass losses. The masses of the eroded grains are distributed according to a power law. This distribution is characterized by minimum ($m_l$) and maximum ($m_u$) grain masses, and the number of grains at each mass within this interval is defined by a mass distribution index ($s$). We assume the refractory silicate grains have a density of $3500\,\text{kg m}^{-3}$, slightly larger than that of Vojáček et al. (2019), since this is similar to grain densities measured for most chondritic meteorites and unmelted silicate micrometeorites (Flynn et al. 2018).

The modelling initiates at an altitude of 180 km, from which the ablation and dynamical equations are integrated for released grains and the main meteoroid every 5 ms using a fourth-order Runge–Kutta method. Grain erosion is treated as quasi-continuous, commencing at a user-specified height, and persisting until the meteoroid is completely consumed or the erosion is manually stopped. Eroded grains undergo ablation with the same ablation coefficient $\sigma$ as the primary parent body. To accommodate potential variations in erosion characteristics, a secondary erosion coefficient $\eta_c$ may be introduced at a height $H_{ec}$ if the light curve or dynamics require it. The integration continues until the meteoroid mass falls below $10^{-14}$ kg and/or the velocity decreases below $3\,\text{km s}^{-1}$. The light curve is computed by summing up the light production of all active fragments at a given time and using the luminous efficiency derived by Vida et al. (2024).

In addition to consideration of the light curve and deceleration of individual fragments, METSIM also models the wake profile. The wake is computed in a 200-m window behind the leading fragment by applying a Gaussian point spread function to all fragments within the window.

The optimization process involves manually adjusting a minimum of thirteen physical parameters to achieve a model fit with the observed data, with more parameters introduced for exceptional events (e.g. flares or sudden disruptions). Table 1 presents the basic model parameters along with their respective physical ranges.

The forward modelling process was conducted manually and iteratively using a GUI that incorporates four main observables: the light curve, the velocity, the deceleration (lag), and wake profiles. The GUI streamlines the fitting procedure by displaying all crucial data and model realizations in real-time and on the same screen. To attain an optimal model solution, a trial-and-error approach is employed to simultaneously fit the observations across all altitudes. The objective is to achieve visually satisfactory fits for each observable. A final solution with small randomly distributed residuals indicates a good quality fit. We chose to require that the deceleration residuals are mostly below 20 m.

**Table 1.** METSIM parameters and parameter ranges used in the forward modelling.

| Parameter | Description | Range |
| --- | --- | --- |
| $v_\infty$ | Meteoroid initial velocity | $12-72\,\text{km s}^{-1}$ |
| $m_\infty$ | Meteoroid initial mass | $10^{-9}-1$ kg |
| $\rho_m$ | Meteoroid bulk density | $100-3500\,\text{kg m}^{-3}$ |
| $\sigma$ | Ablation coefficient | $0.001-0.3\,\text{kg MJ}^{-1}$ |
| $\eta$ | Erosion coefficient | $0.001-2\,\text{kg MJ}^{-1}$ |
| $m_l$ | Smallest grain mass | $10^{-12}-m_u$ kg |
| $m_u$ | Largest grain mass | $m_l - \frac{m_\infty}{2}$ kg |
| $s$ | Grain mass index | $1.4-2.5$ |
| $H_e$ | Erosion start height | $70-120$ km |
| $H_{ec}$ | Erosion change height | $70-120$ km |
| $\eta_c$ | Erosion change coefficient | $0.001-2\,\text{kg MJ}^{-1}$ |
| $\sigma_c$ | Ablation change coefficient | $0.001-0.3\,\text{kg MJ}^{-1}$ |
| $\rho_{mc}$ | Meteoroid bulk density change | $100-3500\,\text{kg m}^{-3}$ |

Evaluating the fit of the wake is challenging as it necessitates replicating its intensity and shape by releasing numerous particles across various mass bins and heights. Computing the wake's brightness involves integrating the contributions of all these particles while accounting for the point spread function of the system perpendicular to the meteor's trajectory. To compare the model's wake with the uncalibrated observed wake, the latter is scaled by normalizing the total area under both brightness curves. However, quantitatively measuring the quality of the wake's fit proves to be a complex task, particularly considering there is a wake for each height. As a result, the fitting procedure initially prioritized obtaining simultaneous fits of the light curve and deceleration profiles. Once an initial fit is achieved, the wake is then examined to improve the overall fit by making smaller changes to secondary parameters.

Finally, we did not find it necessary to adjust the grain mass ranges or size distribution before and after the erosion change. However, for all events we only have good narrow field tracking to provide wake measurements starting just after the change in erosion. We generally find that modest changes in grain mass ranges or distribution only slightly affects the overall light curve and is not detectable in dynamics. Hence we do not expect our modelling to distinguish changes in the grain properties pre- and post-erosion onset.

## 4 RESULTS AND DISCUSSION

The METSIM erosion model was applied to each observed SDA meteor. Fig. 2 shows the fitted model for an SA meteor from 2023 July 31 at 08:15:19 UTC. The light curve is well fit across all altitudes and the lag residuals are all within 20 m. A sequence of figures comparing observations to model fits for the remaining nine events is provided in Appendix A. Tables 2, 3, and 4 present the erosion model parameters derived from the fits.

Our findings suggest that the ablation of SDA meteoroids is characterized by a two-stage erosion process. Most of the data are explained using a relatively low bulk density between 450 and 700 kg m$^{-3}$ for the first part of luminous flight having erosion of refractory grains ranging from 20 to 600 μm until the peak of the light curve. Subsequently, the meteoroid transitions to a second stage dominated by a leading single-body fragment. This suggests a differentially structured meteoroid, where the interior is compact and the crust is composed of weaker material. Investigating the actual chemical composition of these two layers would require additional spectral observations and is beyond the purview of this study.

This bimodal structure is reflected in the dual morphology of the wake, which has a large extent right after erosion change onset and disappears almost completely at the later stage of flight. Accounting for this wake behaviour and the light curve/dynamics requires distinct model densities for each stage. The effective bulk density of the meteoroids is calculated as the mass-weighted average of these values, indicated in the second column of Table 2. The overall average bulk density for all 10 SDA meteors is 1420 ± 100 kg m$^{-3}$. Fig. 3 shows a comparison of our measured bulk density of SDA meteors with those of other major meteor showers. SDA particles have bulk densities which would categorize them with carbonaceous chondrites (Ryabova, Asher & Campbell-Brown 2019; Miyazaki et al. 2023). Together with the QUA and GEM they are in the category of higher bulk densities. The remaining shower meteoroids have low bulk densities consistent with unprocessed cometary material.

In the investigation of a random assortment of mixed sporadic/shower meteors, Vojáček et al. (2019) reported that two-stage erosion was observed in approximately 10 per cent of cases. They specifically highlighted the Draconid and Taurid meteor showers in their sample as exhibiting substantial indications of widespread two-stage erosion. Vida et al. (2024) found the same behaviour for the Orionids. With confidence, we can now include SDA in this category. Furthermore, from Table 3 we see that the initial stage of ablation is characterized by an ablation coefficient comparable to the second stage, while the erosion coefficients are generally different between the two stages. This is in agreement with the results of Vojáček et al. (2019) and Vida et al. (2024). As noted by those authors, this suggests that the material properties are similar for both components but that the packing fraction or porosity differs.

Table 2 shows that for our 1 to 3 mm-sized meteoroids, the fluffy component which ablates first has a porosity near 80 per cent while the leading fragment which dominates later flight and shows little or no wake has roughly half this porosity. The mass is generally equally distributed between the two components.

The original structure of cometary dust at mm-sizes has been best characterized by the *in-situ* measurements of comet 67P/Churyumov-Gerasimenko (67P) by *Rosetta* (Mannel et al. 2019). A major result of the structural analysis of the millimetre and smaller cometary dust observed at 67P is that particles are agglomerations of fundamental grains ranging from nm to micrometre sizes.

Güttler et al. (2019) proposed a unifying classification for cometary dust synthesizing the *in-situ* results from both *Rosetta* and the earlier measurements from Stardust. They suggested that the observations could best explain particle components based on strength, structure, and porosity. Using these properties they identify three dust components: a solid group with high strength and very low porosity, a fluffy group with extremely high porosity and exceptionally low strength, and finally a porous group with intermediate porosities and low strengths. The fluffy and porous groups are complex aggregates made of smaller sub-units while the solid group are more compact and would represent higher temperature condensates such as calcium–aluminium-rich inclusions or chondrules.

For cometary meteoroids in the mm-size range, the best population-level and size estimates are provided by the Grain Impact Analyser and Dust Accumulator (GIADA) instrument on *Rosetta* (Della Corte et al. 2014). From GIADA measurements two populations of dust were identified – compact particles ranging up to mm-sizes and fluffy particles up to 2.5 mm (Fulle et al. 2015). The fluffy category is the same as identified by Güttler et al. (2019) but the compact particles are more likely simply on the low porosity end of their porous group as GIADA cannot separate the porous and solid group.

Most pertinent to our measurements, GIADA detected correlated clusters of many small grains caused by fragmentation of mm-sized fluffy particles passing through the instrument. Interestingly, these fragment clouds were often found to contain a smaller compact component (Fulle et al. 2015; Güttler et al. 2019). We suggest that our SDA measurements are best interpreted similarly, i.e. as a mixture of compact and porous components. Our estimate of the bulk density of the fluffy component is more consistent with the porous dust category. The fact that 90 per cent of our sample in the SDA show this structure indicates that such meteoroids may be common to many comet parent classes.

*Rosetta* results have been interpreted to suggest that hydrocarbons are a major component of cometary particles, making up almost half the volume abundance of mm and larger dust (Fulle et al. 2016). Hydrocarbons may be the organic 'glue' holding the fluffy component together, a concept dating back to the original model of the dustball meteoroid (Hawkes & Jones 1975). If this is the case, these hydrocarbons would contribute little or nothing to the optical light as carbon has no significant light production in our instrument bandpass (Jenniskens 2004) but would be detectable in the meteor deceleration. This might bias our model bulk density estimates making them lower than true values as we would not be accounting for the optically 'hidden' hydrocarbon mass.

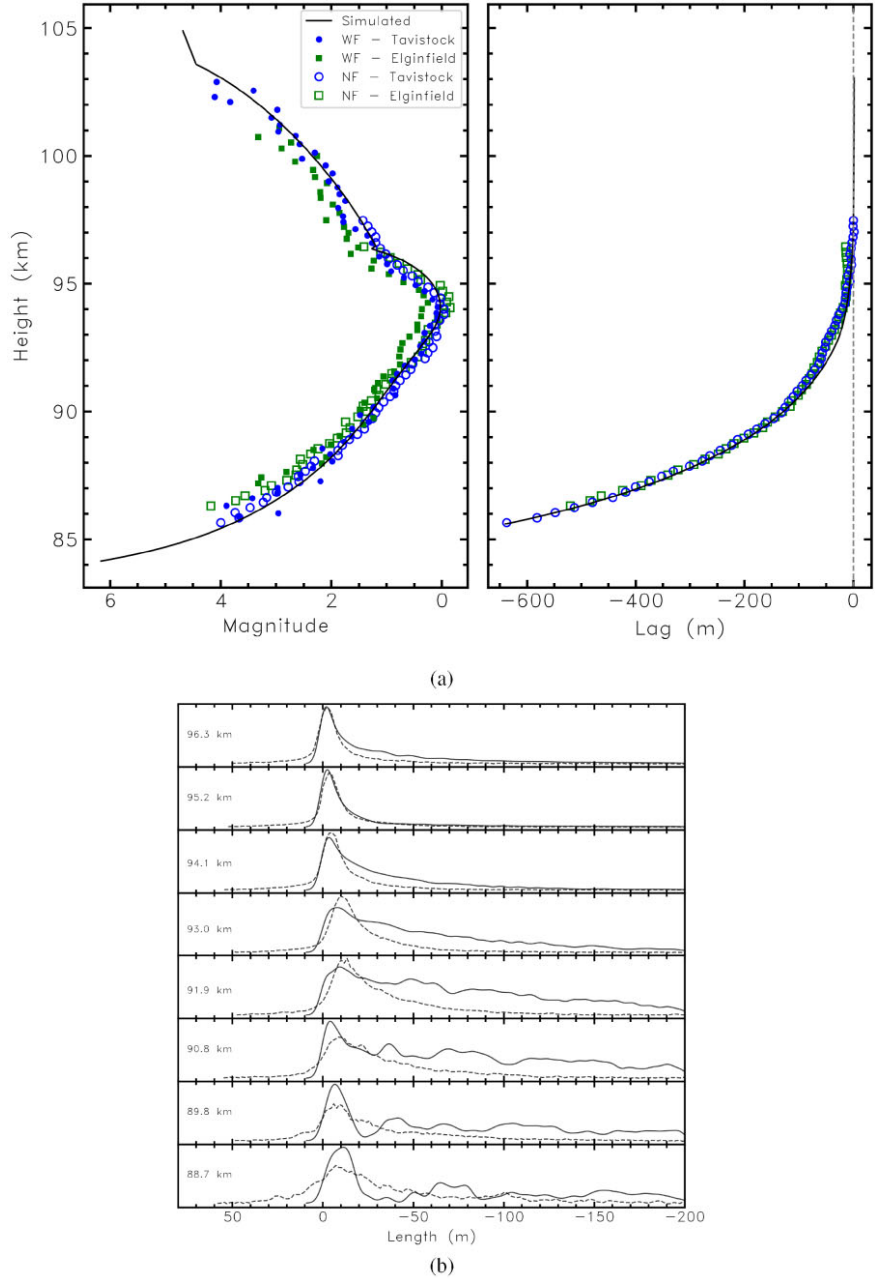

**Figure 2.** (a): Simulated (black) and observed (coloured) light curves and lag profiles for the 2023 July 31 08:15:19 SDA meteor. The widefield and narrowfield observations from the two observing sites are denoted by WF and NF, respectively. (b): Simulated (solid) and observed (dashed) wakes for the 2023 July 31 08:15:19 SDA meteor. The heights of the intensity profiles are inset in the left of each panel.

Results of other studies suggest that our finding of a relatively high bulk density for SDA meteoroids, namely in the carbonaceous chondrite range, is reasonable. Joiret & Koschny (2023) examined 61 SDA meteoroids by analysing their start and end altitudes. The authors deduced meteoroid material characteristics using data from the Canary Islands Long Baseline Observatory (CILBO) system, a dual-station arrangement of image-intensified video cameras. The $K_B$ parameter, a popular proxy metric for the amount of energy required to begin intensive thermal ablation of a meteoroid was introduced by Ceplecha (1967) and applied to their data. The authors found a relatively high $K_B$ of $7.59 \pm 0.13$, a value close to the systematic and analytical errors of analysed GEM meteoroids ($7.34 \pm 0.11$). The $K_B$ value obtained from our analysis is $7.35 \pm 0.07$, which is consistent with the finding of Joiret & Koschny (2023) to within $2\sigma$. Moreover, a spectral investigation by Matlovič et al. (2019) examined meteoroids ranging from millimetre to decimetre sizes, encompassing three SDA meteors with a median $K_B$ value of 7.61. However, no error is given for this median value, presumably because of the limited sample size. Consequently, the variance between our findings and those of Matlovič et al. (2019) is probably attributable to their restricted sample size and/or disparities in bandpass and system sensitivity.

**Table 2.** Bulk parameter values for the best-fitting models of the SDA meteor ensemble. These parameters include the initial mass ($m_\infty$), the bulk density during trajectory segments featuring a visible wake ($\rho_\infty$), the bulk density when the trajectory is governed by a primary fragment ($\rho_s$) and the effective mass loss-weighted average bulk density characterizing the entirety of the meteoroid ($\rho_{\text{eff}}$). Also presented are the porosities associated with the meteoroid material during specific erosion stages: $\phi_\infty$ for the initial erosion, $\phi_s$ when a leading/primary fragment dominates light production in the second erosion stage, and the comprehensive effective porosity $\phi_{\text{eff}}$. The fractional percentage of mass ablated in the first and second stages is denoted by $f_{m_\infty}$ and $f_{m_s}$, respectively. The last column displays the $K_B$ criterion introduced by Ceplecha (1967).

| Date and time (UTC) | $\rho_{\text{eff}}$ (kg m$^{-3}$) | $\phi_{\text{eff}}$ (per cent) | $\rho_\infty$ (kg m$^{-3}$) | $\phi_\infty$ (per cent) | $f_{m_\infty}$ (per cent) | $\rho_s$ (kg m$^{-3}$) | $\phi_s$ (per cent) | $f_{m_s}$ (per cent) | $m_\infty$ (kg) | $K_B$ |
|---|---|---|---|---|---|---|---|---|---|---|
| 2023-07-31 08:15:19 (1) | 1460 | 58.3 | 450 | 87.1 | 34.8 | 2000 | 42.9 | 65.2 | $2.1 \times 10^{-5}$ | 7.10 |
| 2023-07-30 07:42:13 (2) | 1680 | 52 | 600 | 82.9 | 30 | 2150 | 38.6 | 70 | $1.2 \times 10^{-5}$ | 7.40 |
| 2023-07-28 06:08:05 (3) | 1500 | 57.1 | 1500 | 57.1 | 100 | – | – | – | $6.0 \times 10^{-6}$ | 7.27 |
| 2023-07-28 05:45:08 (4) | 1650 | 52.9 | 650 | 81.4 | 45.8 | 2500 | 28.6 | 54.2 | $6.0 \times 10^{-6}$ | 7.07 |
| 2022-08-06 06:40:56 (5) | 820 | 76.7 | 500 | 85.7 | 81.2 | 2200 | 48.6 | 18.8 | $4.60 \times 10^{-6}$ | 7.68 |
| 2022-07-30 08:30:59 (6) | 1210 | 65.4 | 500 | 85.7 | 61.5 | 2350 | 32.9 | 38.5 | $2.3 \times 10^{-6}$ | 7.14 |
| 2022-07-30 07:59:33 (7) | 1700 | 56.3 | 550 | 84.3 | 34.6 | 2300 | 34.3 | 65.4 | $2.6 \times 10^{-6}$ | 7.46 |
| 2020-07-30 07:32:51 (8) | 1410 | 58 | 700 | 80 | 74.5 | 3500 | 0 | 25.5 | $2.1 \times 10^{-6}$ | 7.45 |
| 2020-07-28 06:24:55 (9) | 970 | 72.3 | 450 | 87.1 | 58.7 | 1700 | 51.4 | 41.3 | $5.55 \times 10^{-6}$ | 7.16 |
| 2022-07-29 06:43:41 (10) | 1780 | 52.6 | 1100 | 68.6 | 32 | 2100 | 40 | 68 | $2.25 \times 10^{-6}$ | 7.72 |

**Table 3.** Best-fitting model parameter values for the ensemble of SDA meteors. These include the ablation coefficient, erosion start height, erosion coefficient, largest grain mass, smallest grain mass, grain mass index, erosion change height, erosion change coefficient, ablation change coefficient, and meteoroid bulk density change.

| Identifier | $\sigma$ (kg MJ$^{-1}$) | $H_e$ (km) | $\eta$ (kg MJ$^{-1}$) | $m_u$ (kg) | $m_l$ (kg) | $s$ | $H_{ec}$ (km) | $\eta_{ec}$ (kg MJ$^{-1}$) | $\sigma_c$ (kg MJ$^{-1}$) | $\rho_{mc}$ (kg m$^{-3}$) |
|---|---|---|---|---|---|---|---|---|---|---|
| 1 | 0.02 | 103.7 | 0.038 | $4 \times 10^{-7}$ | $1 \times 10^{-10}$ | 2.02 | 96.5 | 0.6 | 0.018 | 2000 |
| 2 | 0.023 | 100.5 | 0.06 | $6 \times 10^{-7}$ | $1 \times 10^{-10}$ | 1.98 | 97.5 | 0.85 | 0.015 | 2150 |
| 3 | 0.018 | 102.5 | 0.14 | $1.2 \times 10^{-6}$ | $6 \times 10^{-11}$ | 1.85 | – | – | – | – |
| 4 | 0.02 | 105.0 | 0.12 | $1 \times 10^{-7}$ | $5 \times 10^{-11}$ | 2 | 99.3 | 0.001 | 0.008 | 2500 |
| 5 | 0.019 | 96.0 | 0.047 | $2 \times 10^{-8}$ | $7 \times 10^{-11}$ | 1.9 | 91.5 | 0.02 | 0.01 | 2200 |
| 6 | 0.021 | 103.0 | 0.055 | $4 \times 10^{-7}$ | $3 \times 10^{-11}$ | 2.05 | 96.9 | 0.01 | 0.0185 | 2350 |
| 7 | 0.021 | 98.3 | 0.1 | $5 \times 10^{-7}$ | $4 \times 10^{-11}$ | 2.15 | 97.5 | 0.87 | 0.012 | 2300 |
| 8 | 0.022 | 103.0 | 0.11 | $4 \times 10^{-9}$ | $4 \times 10^{-11}$ | 1.9 | 96.9 | 0.04 | 0.014 | 3500 |
| 9 | 0.02 | 97.8 | 0.03 | $2 \times 10^{-7}$ | $1 \times 10^{-10}$ | 1.9 | 94 | 0.019 | 0.014 | 1700 |
| 10 | 0.02 | 99.2 | 0.05 | $2.2 \times 10^{-7}$ | $1 \times 10^{-11}$ | 1.9 | 96.7 | 0.16 | 0.01 | 2100 |

**Table 4.** Peak bolometric absolute magnitudes, pre-atmospheric velocities, and end velocities of the ensemble for the best-fitting models.

| Identifier | $M_{\text{bol}}$ | $v_\infty$ (km s$^{-1}$) | $v_{\text{end}}$ (km s$^{-1}$) |
|---|---|---|---|
| 1 | 0.03 | 42.5 | 15.1 |
| 2 | 0.83 | 42.9 | 3.0 |
| 3 | 1.99 | 38.9 | 3.4 |
| 4 | 1.83 | 43.8 | 3.2 |
| 5 | 1.92 | 40.2 | 3.0 |
| 6 | 2.8 | 42.4 | 14.3 |
| 7 | 2.18 | 41.9 | 3.02 |
| 8 | 2.50 | 42.1 | 8.22 |
| 9 | 2.11 | 43.0 | 3.05 |
| 10 | 2.30 | 42.8 | 3.12 |

It is noteworthy that meteoroids originating from showers characterized by relatively high bulk densities, such as those associated with the SDA, QUA, and GEM showers exhibit a tendency towards orbits with low perihelia. Fig. 4 shows the perihelia distances of the showers in Fig. 3. The showers with densities typical of carbonaceous chondrites have perihelia below 0.15 au, while all others have perihelia above 0.6 au. Orbital simulations indicate that approximately 1500 yr ago the orbit of QUA meteoroids approached the Sun to within 0.10 au, and has since moved out to just below 1 au due to frequent close encounters with Jupiter (Porubčan & Kornoš 2005). Such close perihelion distances result in high temperatures that can lead to the desiccation of hydrated minerals and the loss of other volatile elements such as sodium (Koten et al. 2006).

Borovička et al. (2005) found that small sporadic meteoroids with perihelia less than 0.2 au were completely depleted in Na and compacted. The Geminids have also been shown to be highly depleted in Na and other volatile elements (Borovička 2010). At 0.14 au, GEM meteoroids are heated by solar radiation to around 700 K. The Fe/Mg ratio is also similar to Halley-type cometary material, implying 3200 Phaethon is a remnant cometary nucleus (Borovička 2007). Similarly, SDA meteoroids were found to be Na-free and compacted with temperatures at perihelion reaching 1000 K (Borovička 2010; Matlovič et al. 2019). Alongside the stream's considerable age, estimated to be between 10 000 to 20 000 yr (Abedin et al. 2018), it is highly likely that the majority of SDAs have undergone the loss of their sodium content through thermal desorption. It is worth noting that a few SDA meteor spectra identify Na lines (Matlovič et al. 2015, 2019), suggesting the possibility of a semicontinuous production of meteoroids. This implies varying ages of close exposure to the Sun for individual meteoroids within the stream. Overall, the high densities observed in these three showers may be attributed to the small perihelia of their orbits.

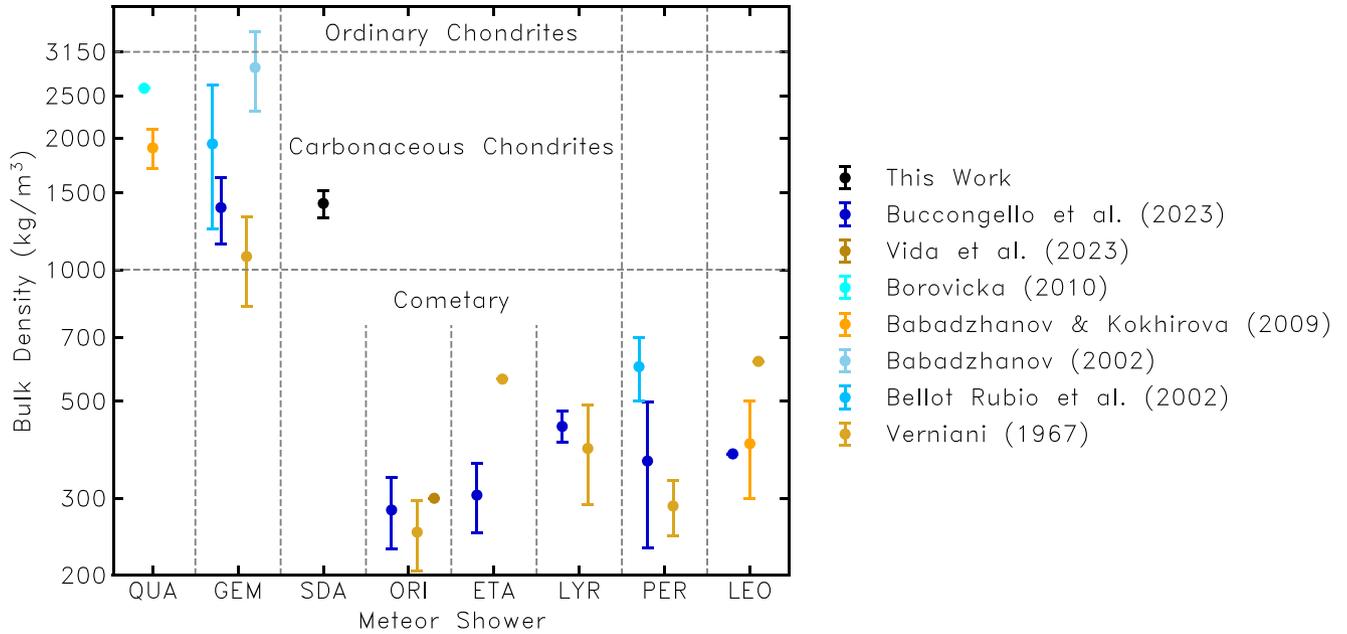

**Figure 3.** Average bulk densities of primary shower meteoroids. SDA particulates are found to have densities in the carbonaceous chondrite class. Estimates for other showers are obtained from Buccongello et al. (2024), Vida et al. (2024), Borovička (2010), Babadzhanov & Kokhirova (2009), Babadzhanov (2002), Bellot Rubio et al. (2002), and Verniani (1967).

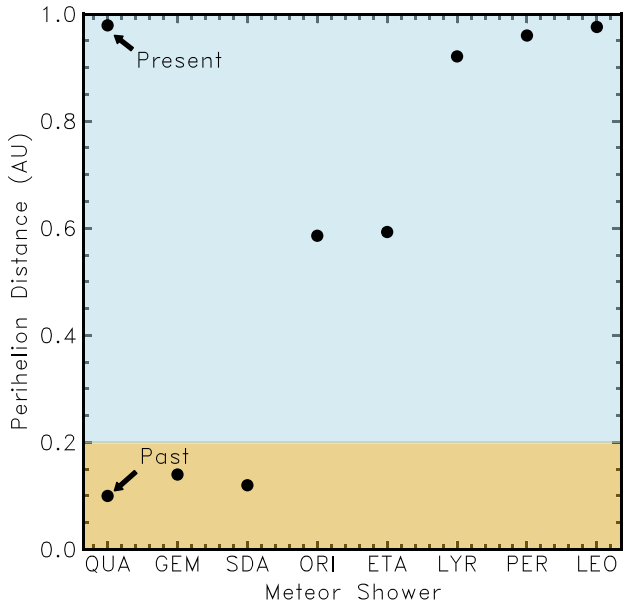

**Figure 4.** Perihelia distances of active meteor showers. Showers with the highest bulk densities approach the Sun most closely. The boundary in shaded areas signifies the distance at which sodium sublimation starts to play a significant role, as detailed by Borovička et al. (2005).

## 5 CONCLUSIONS

Using precise optical observations obtained through CAMO and applying the METSIM meteoroid erosion modelling tool, we estimated the physical properties of the SDA shower. We find that the average bulk density for ten SDA meteoroids in the 1–3 mm diameter range to be $1420 \pm 100$ kg m$^{-3}$ placing them in the carbonaceous chondrite class.

The results of the forward modelling process reveal that almost all the SDA meteoroids undergo a two-stage erosion process. The initial stage is characterized by relatively low bulk density ($700 \pm 110$ kg m$^{-3}$) while the later stage is dominated by a high density ($2310 \pm 160$ kg m$^{-3}$) leading single-body fragment. The mass removed in each stage is almost equal and the ablation coefficient shows no significant difference before and after erosion change. The erosion coefficient shows a significant change (almost an order of magnitude) between the erosion changes on average, but not in a consistent sense, with some events eroding much more vigorously before the erosion change and vice versa.

Interpreting these results in the context of recent Rosetta findings suggests that the SDA are equal conglomerates of a low-density, porous component, and an embedded compact component. The similar ablation coefficients for each component suggest that they are made from the same fundamental grains, a result also found by Vojáček et al. (2019). The difference in erosion coefficients suggests that there is more randomness to the cohesiveness of the grains in the two components.

Comparison with other major meteor showers illustrates that SDA meteoroids share physical similarities with showers having low perihelion distances such as the QUA – which are potentially genetically related to the SDA as described by Abedin et al. (2018) – and GEM. All three are characterized by high bulk densities. Despite the overall success of the modelling approach, some limitations are apparent. Notably there remain cases where the wake profiles are not perfectly matched, indicating the model's simplicity compared to the complexity of the fragmentation process. Additionally, the components of the meteoroids may not all produce light in the same manner which could lead to biases in our estimation of physical properties. Finally, many of our events have initial erosion onset (as estimated by the model) prior to detection which restricts the precision of ablation coefficient estimates in the early stage of flight. Similarly, the finite time required for the narrow field cameras to begin tracking means we have limited wake estimates prior to the change in erosion, further restricting our inversion constraints.

**ACKNOWLEDGEMENTS**

Funding for this work was provided through the NASA Meteoroid Environment Office cooperative agreement 80NSSC21M0073 and NASA Contract 80MSFC18C0011. We thank B. Cooke for suggestions which have improved the manuscript.

**DATA AVAILABILITY**

The data underlying this article will be shared on reasonable request to the corresponding author.

**APPENDIX: MODEL FITS**

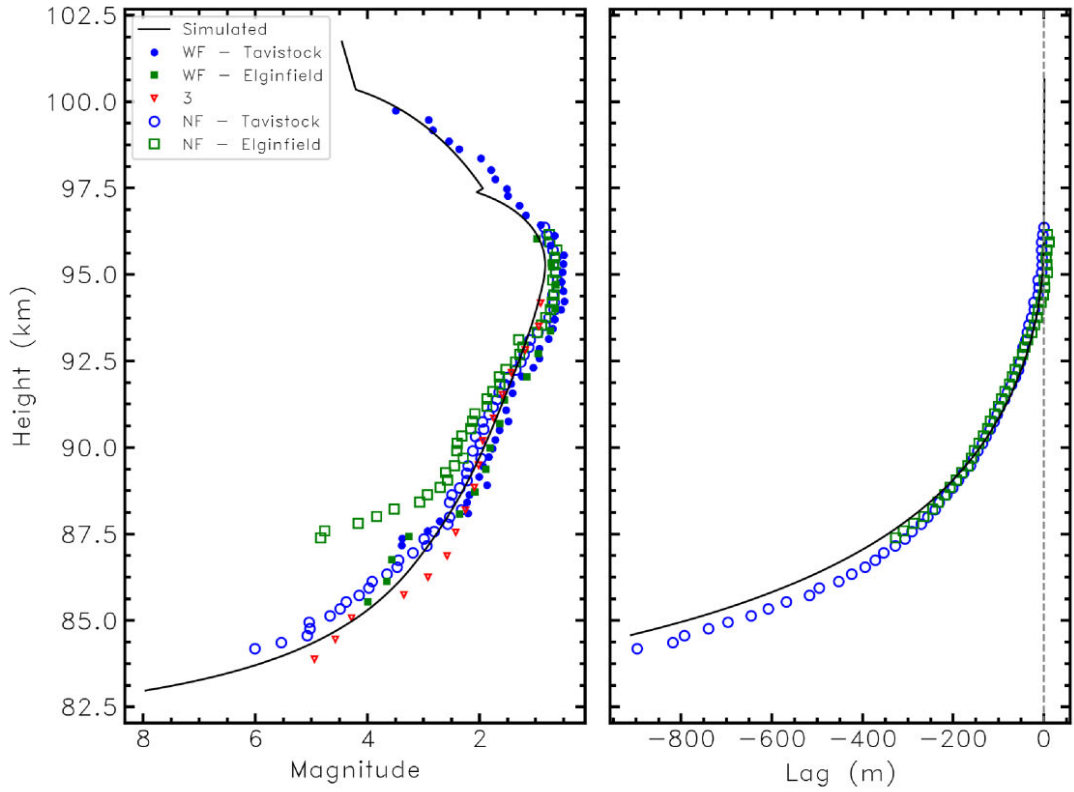

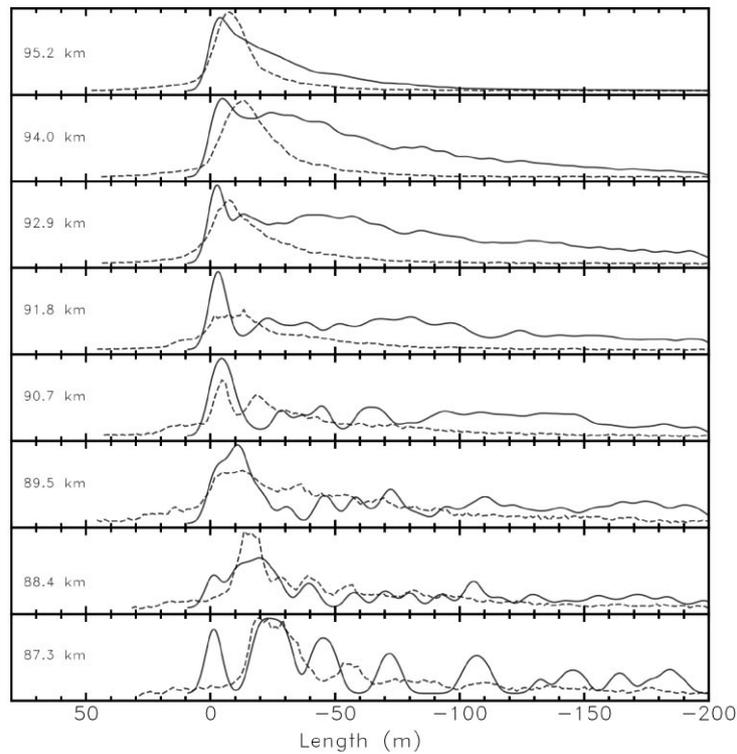

**Figure A1.** (a): Simulated (black) and observed (coloured) light curves and lag profiles for the 2023 July 30 07:42:13 SDA meteor. The widefield and narrowfield observations from the two observing sites are denoted by WF and NF, respectively. (b): Simulated (solid) and observed (dashed) wakes for the 2023 July 30 07:42:13 SDA meteor. The heights of the intensity profiles are inset in the left of each panel.

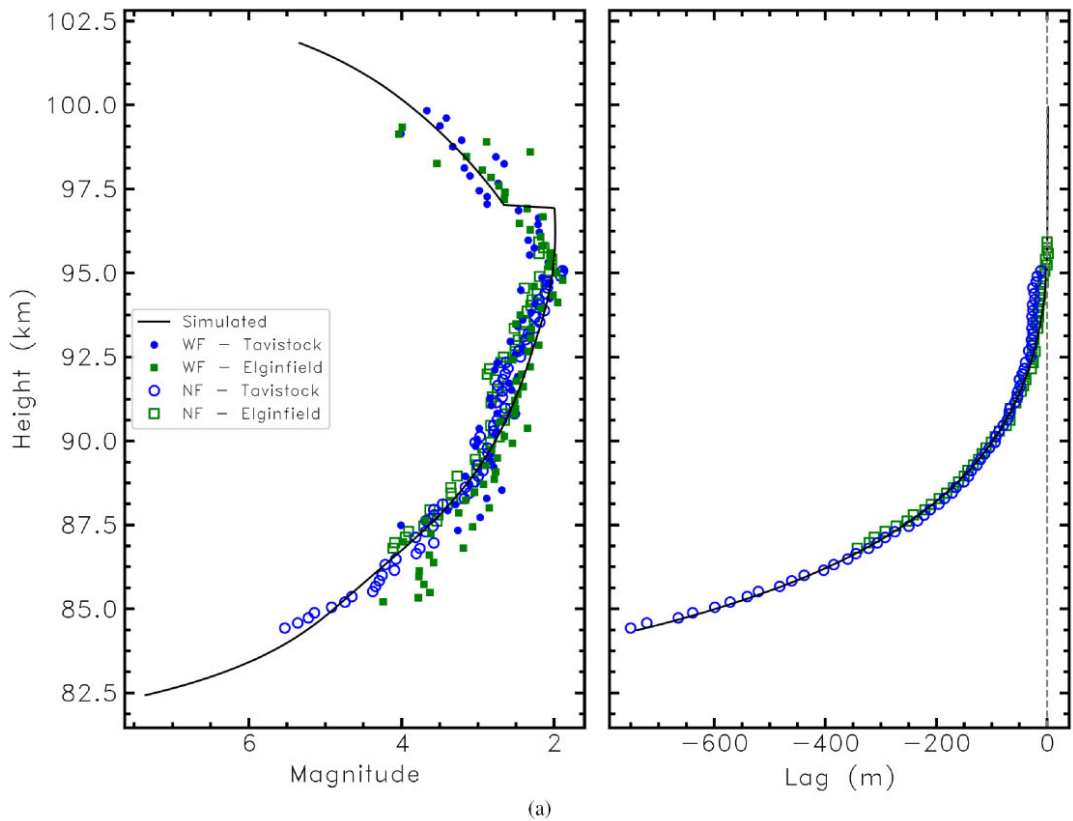

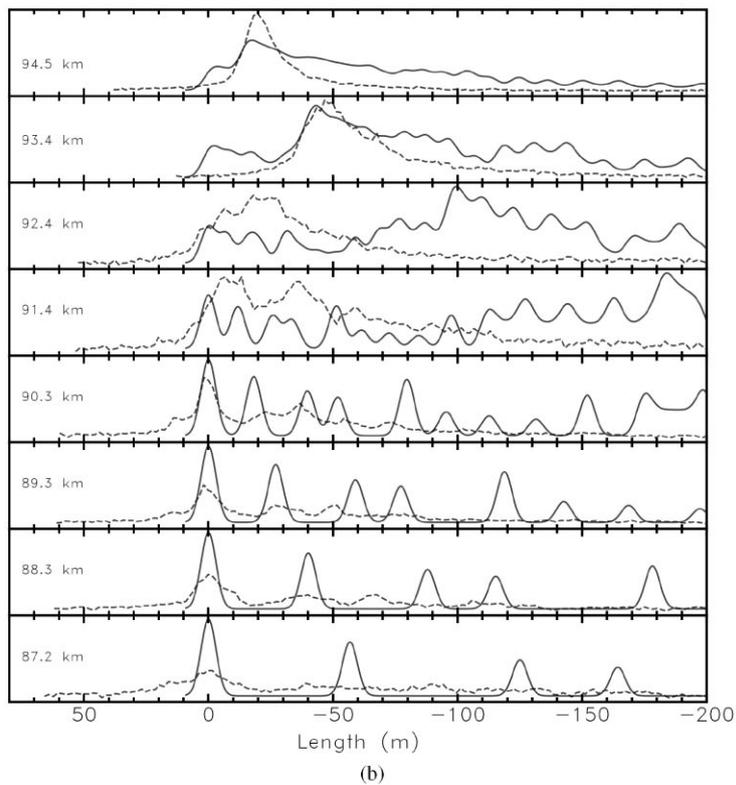

**Figure A2.** (a): Simulated (black) and observed (coloured) light curves and lag profiles for the 2023 July 28 06:08:05 SDA meteor. The widefield and narrowfield observations from the two observing sites are denoted by WF and NF, respectively. (b): Simulated (solid) and observed (dashed) wakes for the 2023 July 28 06:08:05 SDA meteor. The heights of the intensity profiles are inset in the left of each panel.

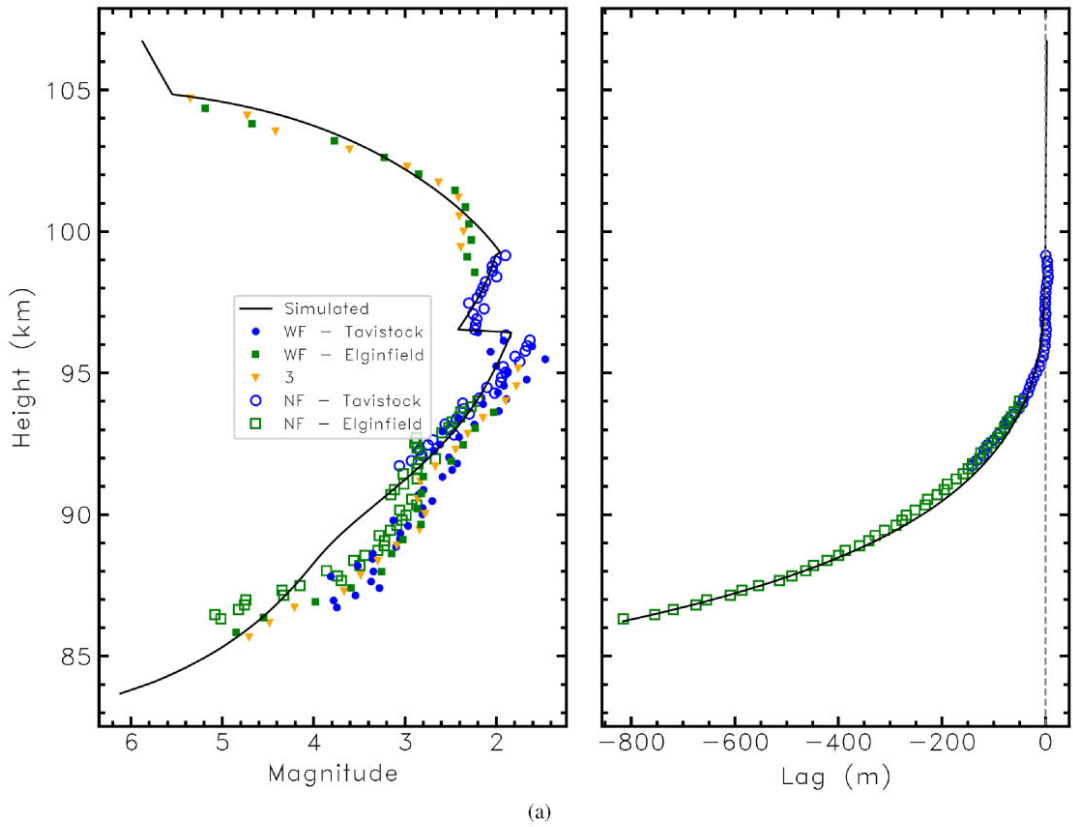

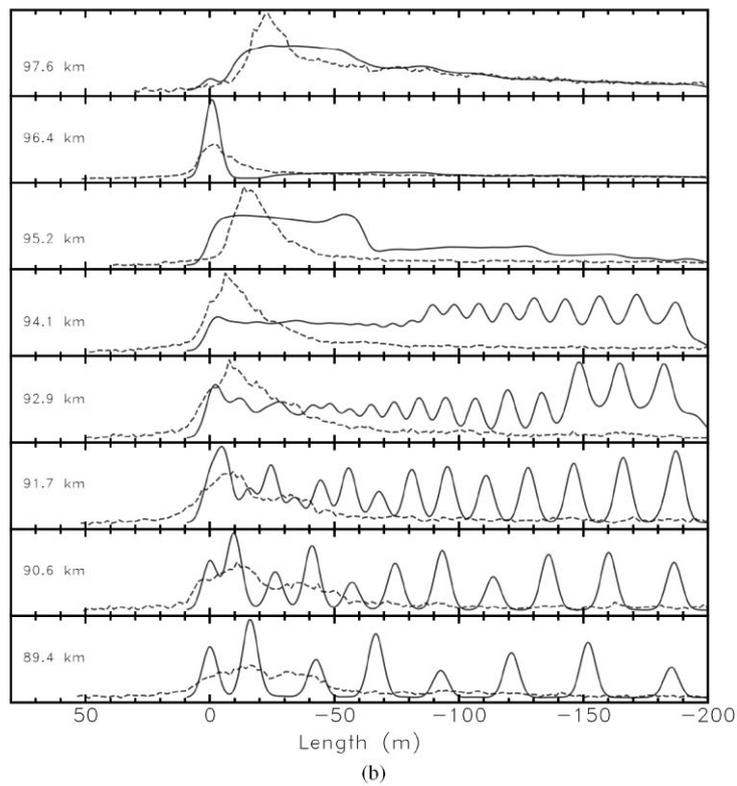

**Figure A3.** (a): Simulated (black) and observed (coloured) light curves and lag profiles for the 2023 July 28 05:45:08 SDA meteor. The widefield and narrowfield observations from the two observing sites are denoted by WF and NF, respectively. (b): Simulated (solid) and observed (dashed) wakes for the 2023 July 28 05:45:08 SDA meteor. The heights of the intensity profiles are inset in the left of each panel.

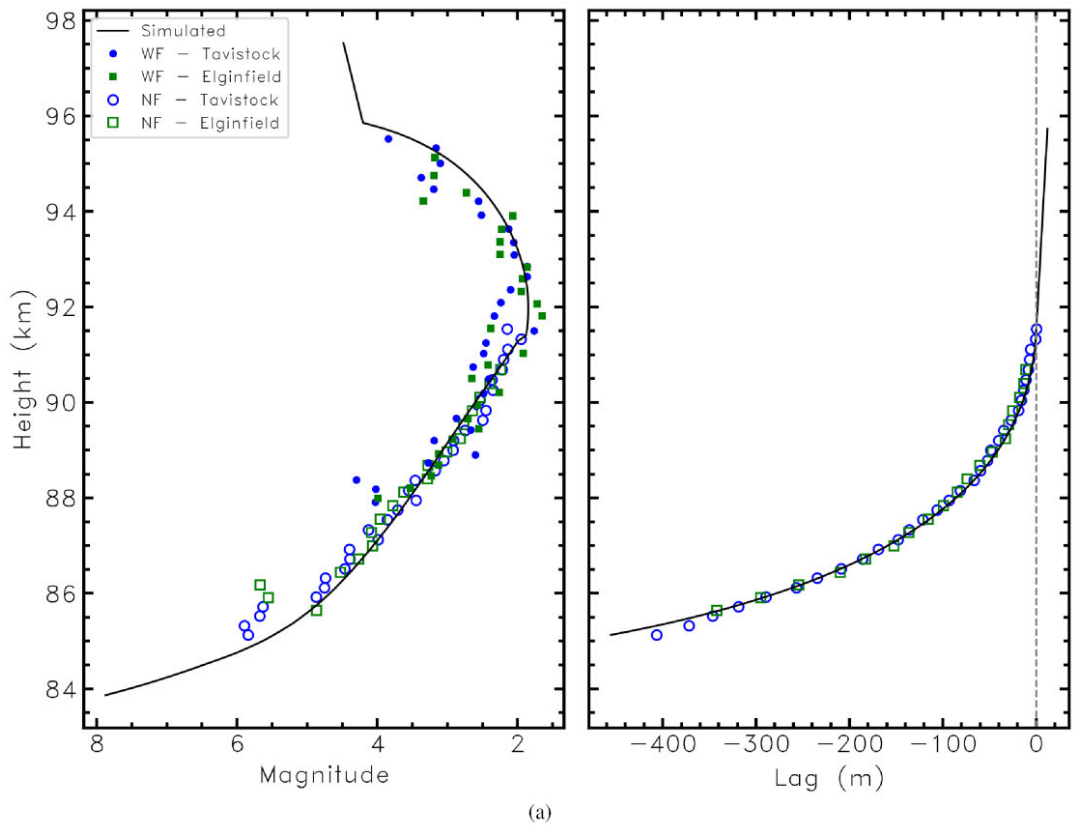

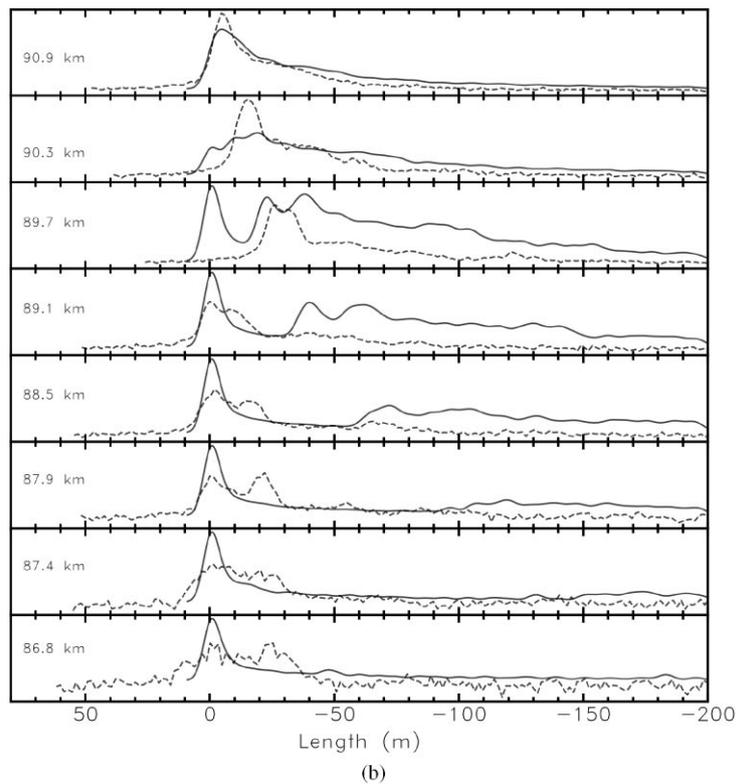

**Figure A4.** (a): Simulated (black) and observed (coloured) light curves and lag profiles for the 2022 August 6 06:40:56 SDA meteor. The widefield and narrowfield observations from the two observing sites are denoted by WF and NF, respectively. (b): Simulated (solid) and observed (dashed) wakes for the 2022 August 6 06:40:56 SDA meteor. The heights of the intensity profiles are inset in the left of each panel.

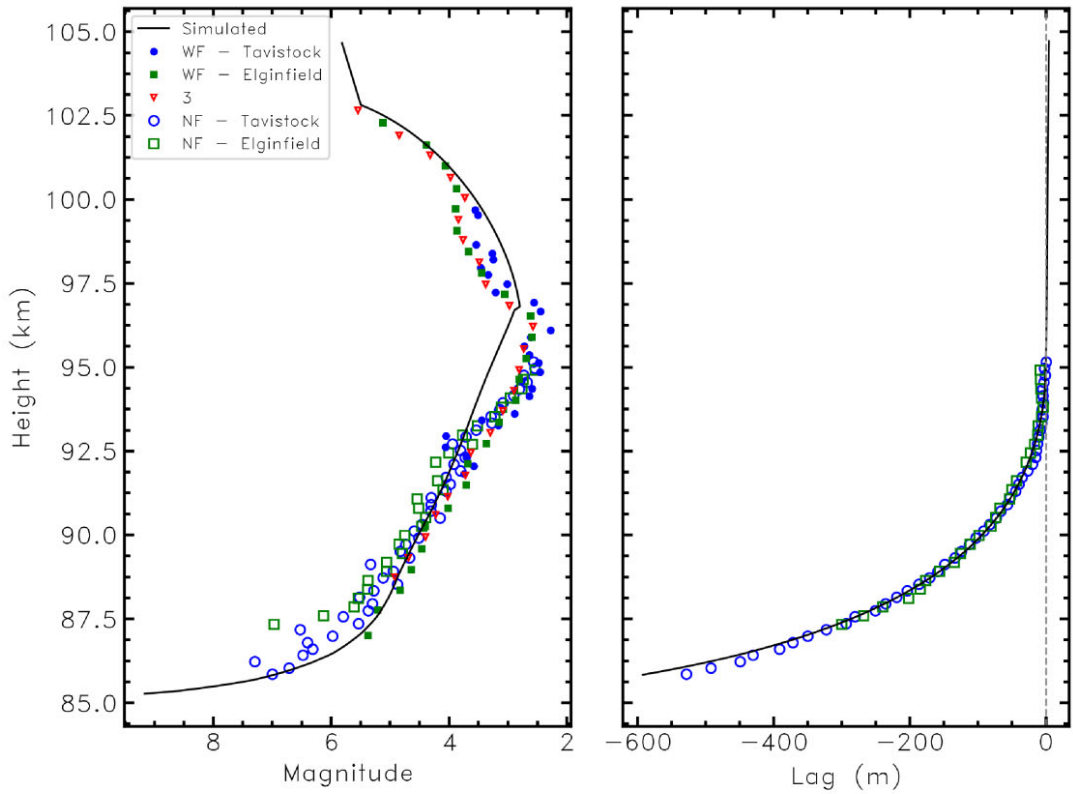

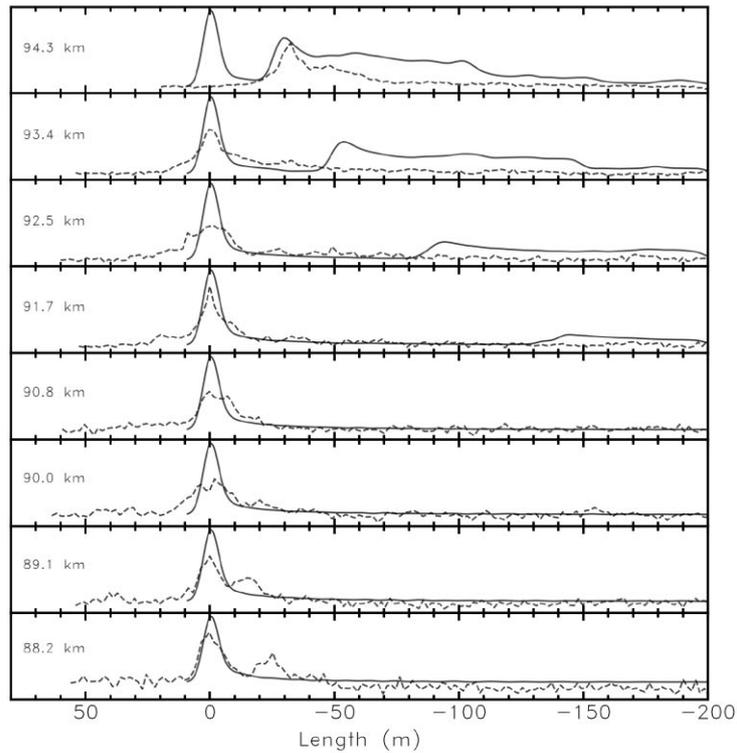

**Figure A5.** (a): Simulated (black) and observed (coloured) light curves and lag profiles for the 2022 July 30 08:30:59 SDA meteor. The widefield and narrowfield observations from the two observing sites are denoted by WF and NF, respectively. (b): Simulated (solid) and observed (dashed) wakes for the 2022 July 30 08:30:59 SDA meteor. The heights of the intensity profiles are inset in the left of each panel.

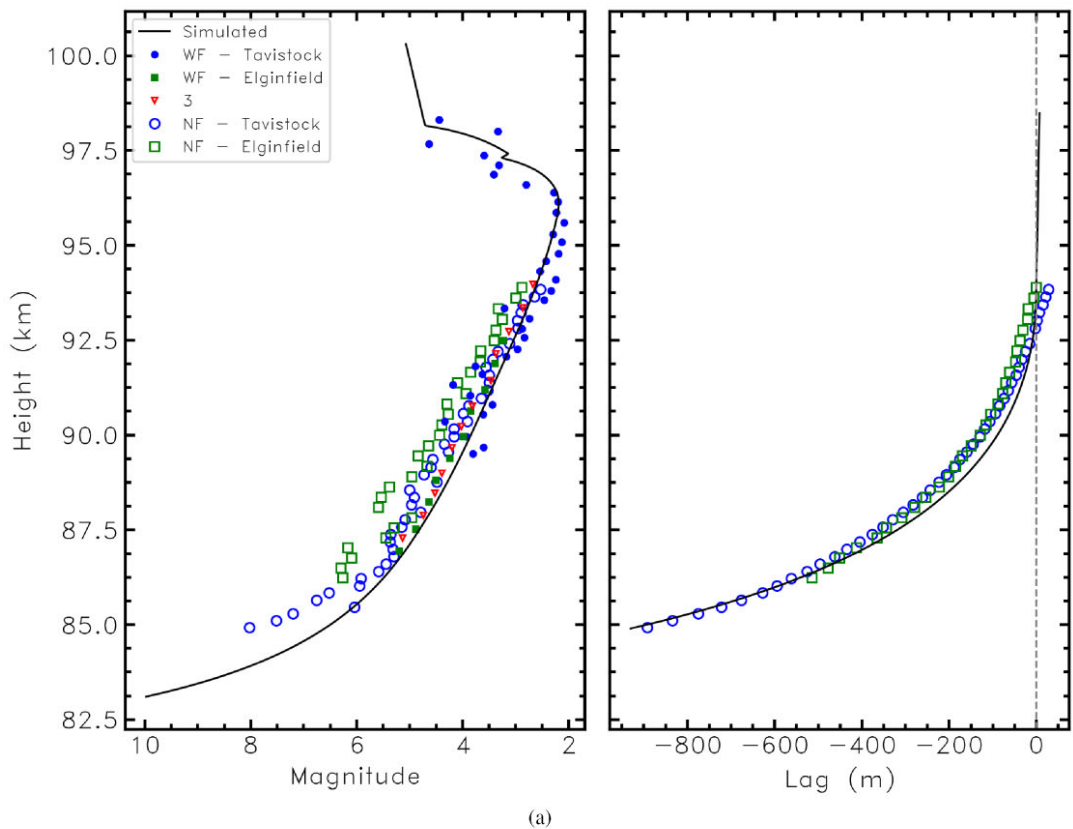

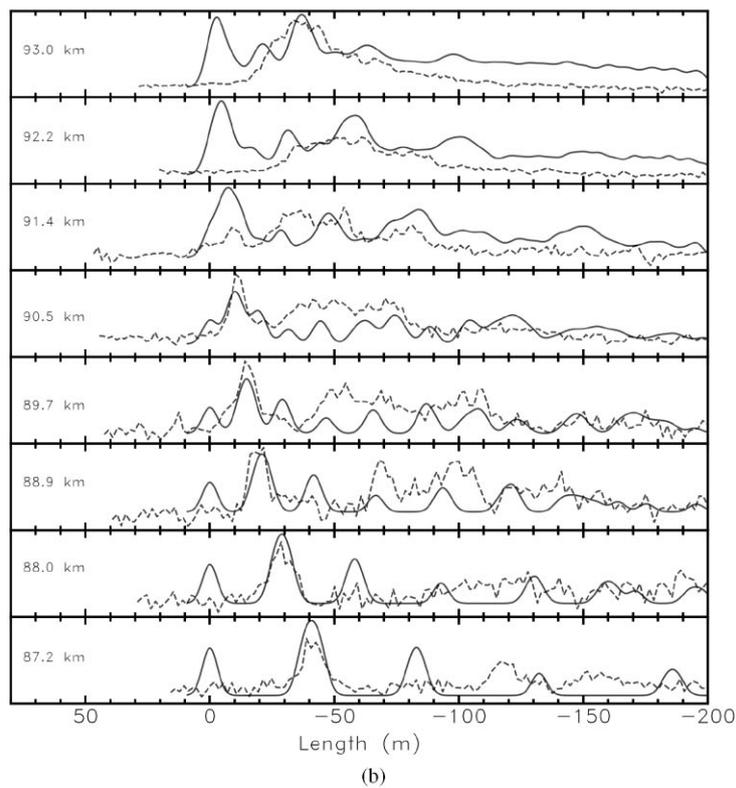

**Figure A6.** (a): Simulated (black) and observed (coloured) light curves and lag profiles for the 2022 July 30 07:59:33 SDA meteor. The widefield and narrowfield observations from the two observing sites are denoted by WF and NF, respectively. (b): Simulated (solid) and observed (dashed) wakes for the 2022 July 30 07:59:33 SDA meteor. The heights of the intensity profiles are inset in the left of each panel.

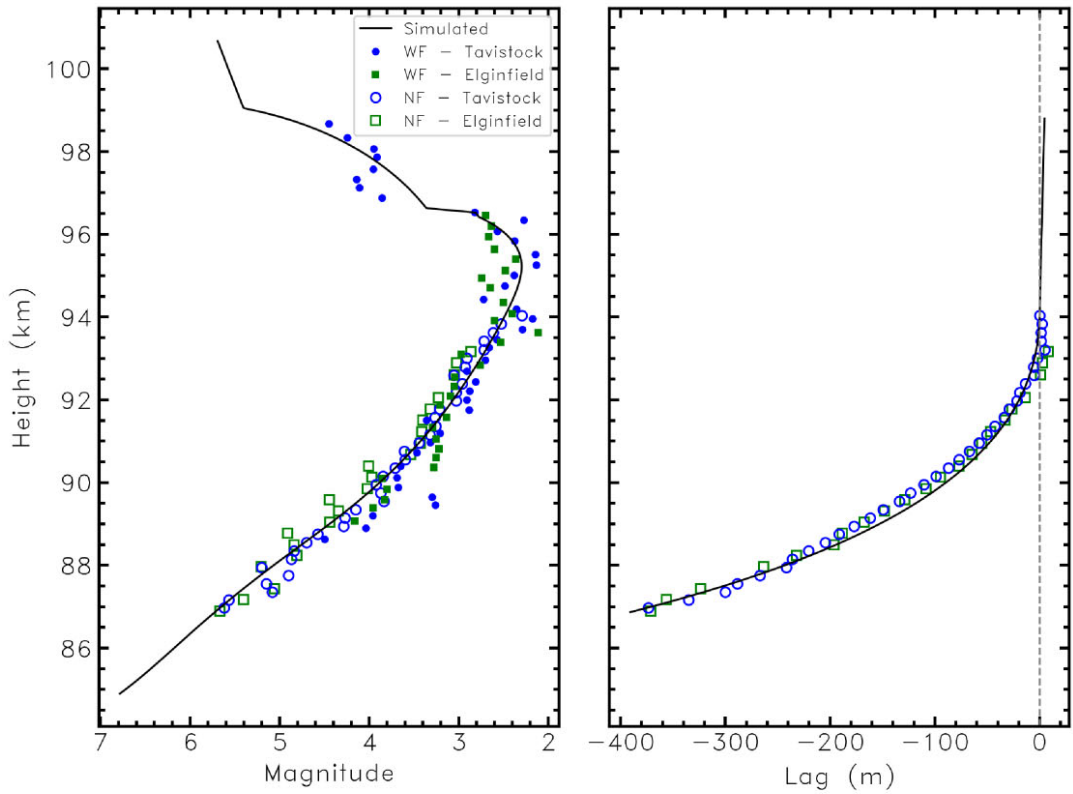

(a)

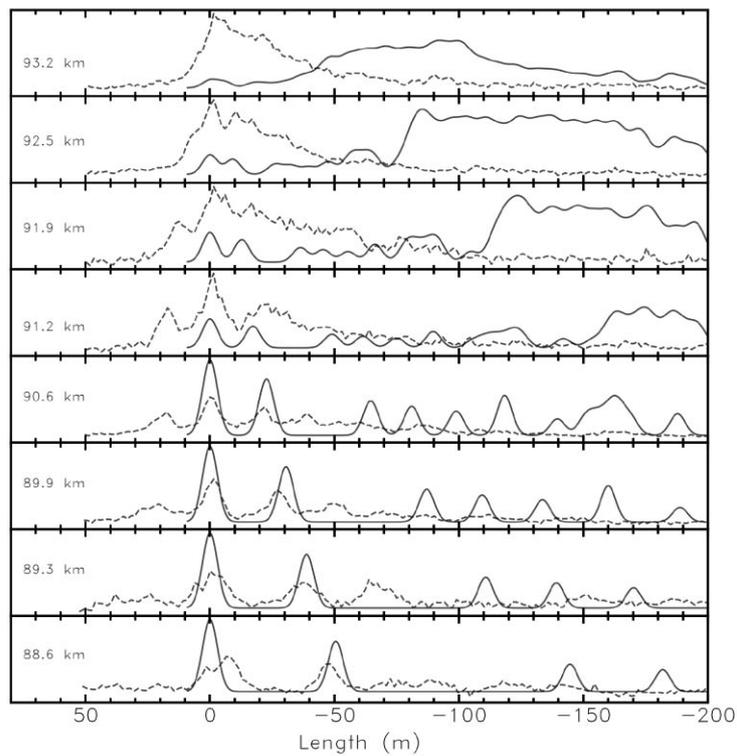

(b)

**Figure A7.** (a): Simulated (black) and observed (coloured) light curves and lag profiles for the 2022 July 29 06:43:41 SDA meteor. The widefield and narrowfield observations from the two observing sites are denoted by WF and NF, respectively. (b): Simulated (solid) and observed (dashed) wakes for the 2022 July 29 06:43:41 SDA meteor. The heights of the intensity profiles are inset in the left of each panel.

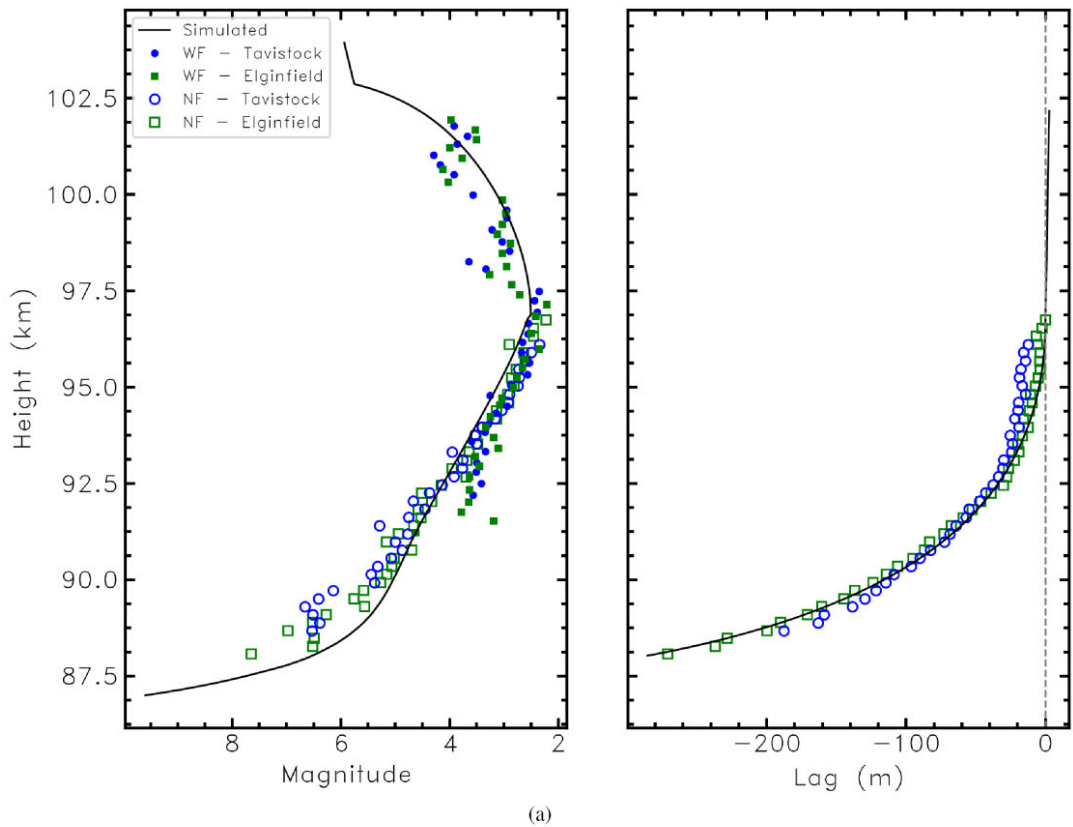

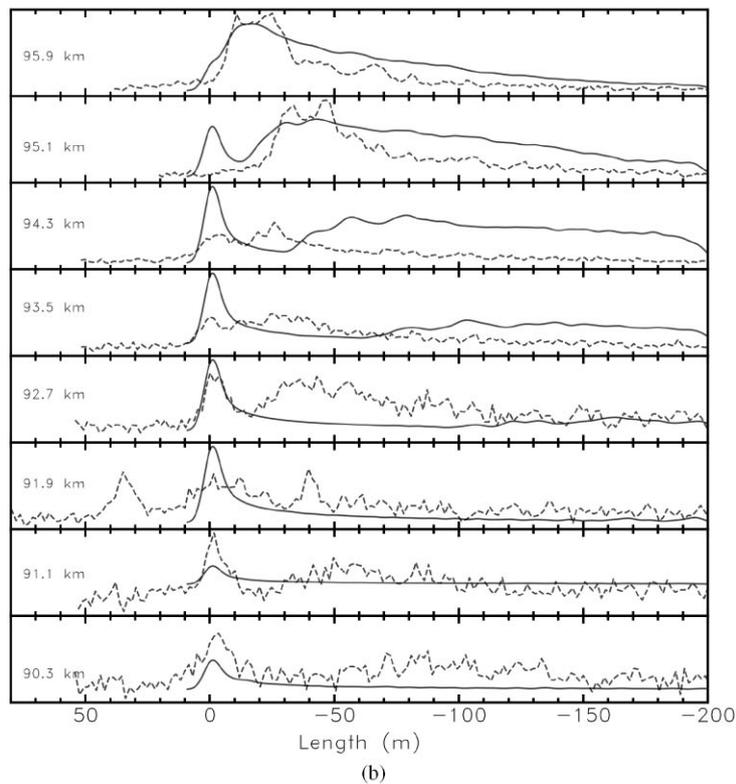

**Figure A8.** (a): Simulated (black) and observed (coloured) light curves and lag profiles for the 2020 July 30 07:32:51 SDA meteor. The widefield and narrowfield observations from the two observing sites are denoted by WF and NF, respectively. (b): Simulated (solid) and observed (dashed) wakes for the 2020 July 30 07:32:51 SDA meteor. The heights of the intensity profiles are inset in the left of each panel.

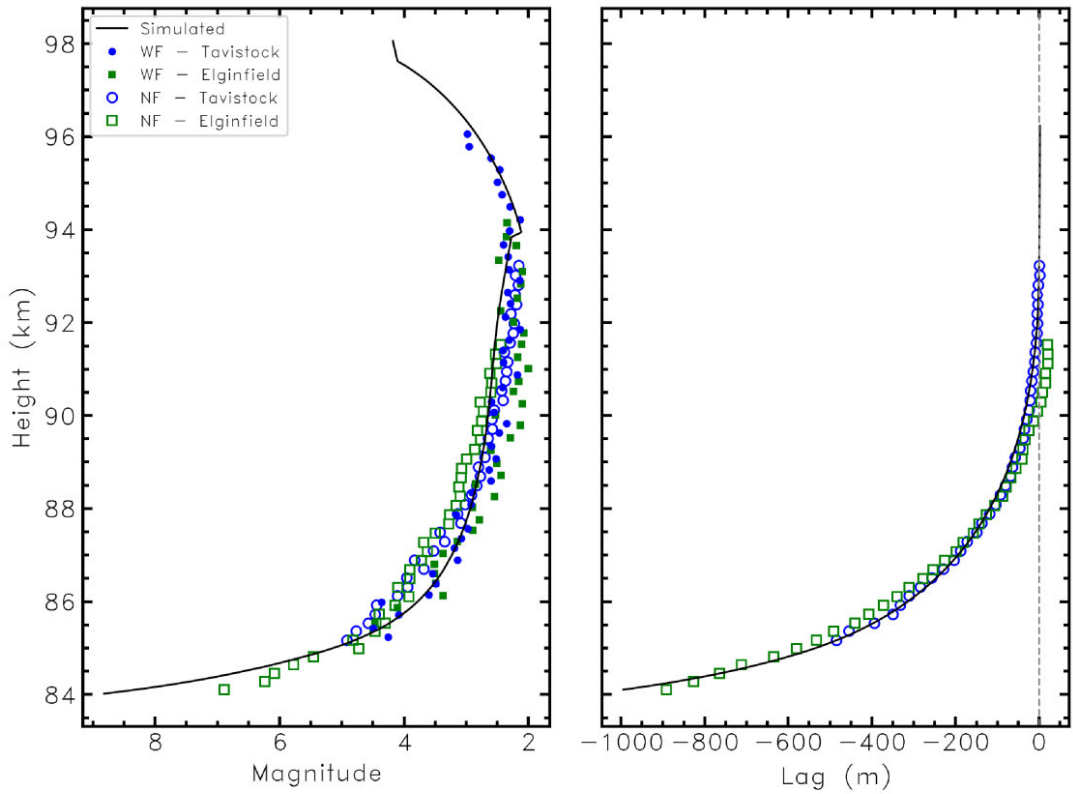

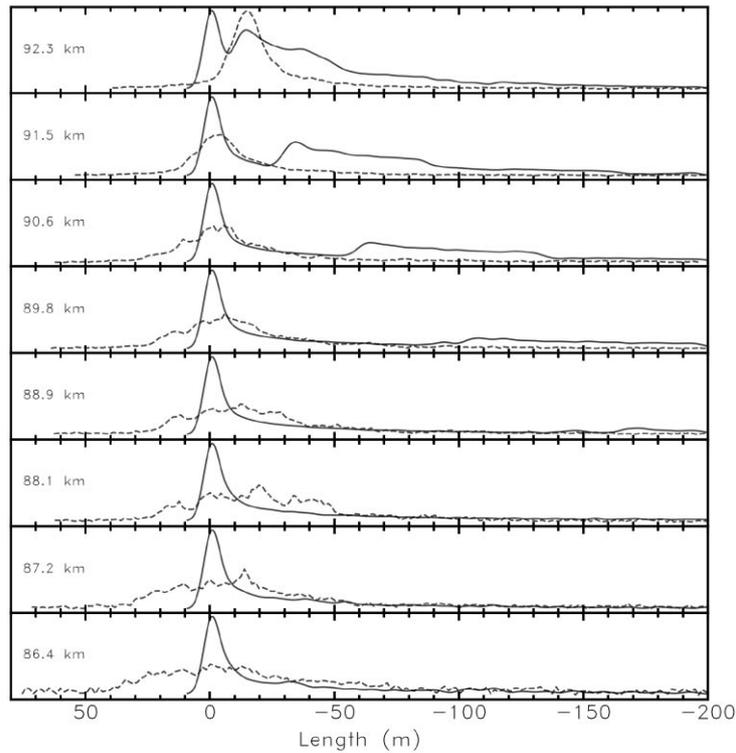

**Figure A9.** (a): Simulated (black) and observed (coloured) light curves and lag profiles for the 2020 July 28 06:24:55 SDA meteor. The widefield and narrowfield observations from the two observing sites are denoted by WF and NF, respectively. (b): Simulated (solid) and observed (dashed) wakes for the 2020 July 28 06:24:55 SDA meteor. The heights of the intensity profiles are inset in the left of each panel.

This paper has been typeset from a TeX/LaTeX file prepared by the author.